\documentstyle[11pt]{article}

\def\Bagd{B_\al\,^{\ga\de}}
\def\eth{e_\al\,^\vartheta}
\def\Habg{H_{\al\be}\,^\ga}

\def\Rabgd{R_{\al\be}\,^{\ga\de}}
\def\Rab{R_{\al\be}}
\def\Tabg{T_{\al\be}\,^\ga}

\def\ar{\rightarrow}

\def\bib{\bibitem}

\def\dem{\det e^{-1}}

\def\intf{\int d^{4}x\,}

\def\lar{\longrightarrow}
\def\lbr{\lbrack}

\def\pa{\partial}

\def\rbr{\rbrack}

\def\al{\alpha}
\def\be{\beta}
\def\ga{\gamma}
\def\Ga{\it\Gamma}
\def\de{\delta}
\def\ep{\varepsilon}
\def\ze{\zeta}
\def\th{\vartheta}
\def\ka{\kappa\,}

\def\La{{\it\Lambda}}

\def\Si{\Sigma}
\def\va{\varphi}
\def\om{\omega}

\def\beq{\begin{equation}}
\def\eeq{\end{equation}}
\def\bed{\begin{displaymath}}
\def\eed{\end{displaymath}}
\def\beqq{\begin{eqnarray}}
\def\eeqq{\end{eqnarray}}
\def\bedd{\begin{eqnarray*}}
\def\eedd{\end{eqnarray*}}

\title{\huge{\bf{Local Lorentz invariance and a new theory of gravitation equivalent to General Relativity}}}
\author{\vspace{0.5cm}\\
C. Wiesendanger\\
Aurorastr. 24\\
8032 Zurich, Switzerland\\
{\it christian.wiesendanger@ubs.com}}
\date{{\it Revised} January 1, 2019}

\begin{document}

\maketitle

\begin{abstract}
A gauge theory of the Lorentz group with a mass-dimension one gauge field coupling to matter of any spin is developed. As a completely new feature the "Vierbein" assuring local gauge invariance enters not as an independent dynamical field, but emerges as a functional of the Lorentz gauge field. The underlying geometry of the theory turns out to be a $SO(1,3)$ Banach bundle. The most general action which is re- normalizable by power-counting is constructed in terms of the gauge field and its first derivatives. It contains no higher derivative terms in the gauge field which destroy unitarity in the usual renormalizable $R^2$-theories of gravitation. Finally equivalence of the Lorentz gauge field theory coupled to spin zero matter with General Relativity is established.\\
\end{abstract}

\clearpage

\section{Introduction}

\paragraph{}
Gravity has defied so far all attempts at consistent quantization. In fact, Einstein's  theory of General Relativity (GR) and its generalizations turn out to be either not renormalizable or do not respect unitarity at the quantum level. Let us revisit the key reasons for that.

In the case of GR as defined by the Einstein-Hilbert (EH) action $S_{EH} = \frac{1}{\ka}\, \int \sqrt{- g} \, R$ simple power-counting in the fundamental dimension-zero field $g$, the metric, or equivalently $e$, the Vierbein, and the fact that the coupling constant $\ka$ carries mass-dimension minus two allow to demonstrate that the loop expansion of the quantum effective action contains divergent contributions which would have to be cancelled by counterterms of ever higher mass-dimension - destroying renormalizability (note that dimensions are counted in powers of mass as usual) \cite{buc}.

Various unsuccessful attempts at resolving this fundamental problem have been made.

A first important one is to improve the ultraviolet behaviour of loop integrals by adding $R^2$-terms to the EH-action. This makes the theory technically renormalizable, however at the price of adding non-physical ghosts which do not decouple from the physical sector of the theory spanned by quantum states with positive energy and positive norm \cite{stel}. These ghosts have their origin in the $R^2$-terms containing second derivatives in the fundamental fields $g$ or $e$ in addition to the fields and their first derivatives occurring in the action. As a result the quantized theory violates unitarity.

Another approach aims at establishing gravity as a gauge theory similar to Yang-Mills theories - a comprehensive overview is given e.g. in \cite{hehl1, hehl2}. The gauge groups considered are the translation group $T^4$ with the Vierbein $e$ as gauge field \cite{cho1, wie1} or the Poincar$\acute{\mbox{e}}$ group $P^4$ with the Vierbein $e$ and a general affine connection $\Ga$ carrying curvature and torsion as gauge fields \cite{cho2, hay, wie2}. Considering the homogenous Lorentz group {\bf SO(1,3)\/} as the gauge group, local translations linked to the spacetime part of the local Lorentz transformations are coming in through the back door in the existing approaches, making them equivalent to gauging the Poincar$\acute{\mbox{e}}$ group \cite{uti, kib}. In addition, there are more general geometric approaches to local gauge invariance defined on higher-dimensional spaces linking gravity with gauge fields related to inner degrees of freedom \cite{chang}. In all of these theories spacetime as the base space of the corresponding bundle becomes curved and "interlocked" with the fibres of the respective bundle \cite{chang}.

And in all cases one ends up with a non-renormalizable or a non-unitary theory depending on what specific action is proposed to describe the field dynamics. The fundamental problem goes back to the dimension-zero field $g$ or $e$ being a seemingly indispensable dynamical field in the theory - or is it? It certainly is when gauging the translation or the Poincar$\acute{\mbox{e}}$ group where $e$ necessarily enters as a dynamical gauge field, in the latter case together with the dimension-one connection $\Ga$ as a complementary gauge field.

But when carefully gauging the homogenous Lorentz group should there not just be one gauge field, the dimension-one connection $\Ga$? And if a kind of Vierbein has to enter the theory should it not rather emerge as a functional of the independent gauge field, i.e. the connection, than as an independent field to which by construction no gauge degrees of freedom correspond? And if the only dynamical field is of dimension one should there not be actions for such a theory which are renormalizable at least by power-counting?

We are not aware of positive answers to the questions above in the litera- ture. So in this paper we develop a gauge field theory of the Lorentz group {\bf SO(1,3)\/} which a) is new as the only dynamical field is the dimension-one Lorentz gauge field in terms of which all else can be expressed and which allows for actions renormalizable by power-counting, yet without higher derivative terms; and b) is equivalent to GR in a limiting case.

To do so we recast in the next section global Lorentz symmetry as an inner symmetry \cite{wie2}. In the third section gauging the Lorentz group {\bf SO(1,3)\/} introduces a covariant derivative with a Lorentz gauge field whose transformation behaviour is established under local gauge transformations. In this step a functional of the Lorentz gauge field emerges which plays the role of a Vierbein. In section four we determine the covariant field strength tensors of the theory and in section five we establish the underlying geometric structure of the theory. In section six we determine the most general invariant gauge field action of mass-dimension $\leq 4$ in the gauge field and its first derivatives which is renormalizable by power-counting. The final two sections are devoted to establish the aforementioned equivalence to GR and to resolve the puzzle why a non-renormalizable theory like GR given in terms of a dimension-zero gauge field can be the limiting case of a renormalizable one expressed in terms of a dimension-one gauge field.

Throughout this paper we work on Minkowski spacetime {\bf M\/}$^{4}$ $\equiv$ ({\bf R\/}$^{4}$,$\eta$) with Cartesian coordinates. $\eta=\mbox{diag}(-1,1,1,1)$ is the flat spacetime metric. Indices $\al,\be,\ga,...$ denote quantities defined on {\bf M\/}$^{4}$ which transform covariantly w.r.t. the global Lorentz group. They are correspondingly raised and lowered with $\eta$.

\section{Global Lorentz invariance as an inner symmetry}

\paragraph{}
In this section we revisit the invariance of field theories under Lorentz and spacetime translations treated as global inner symmetry transformations.

Let us start with a field $\va(x)$ which lives in a given representation of the Lorentz group {\bf SO(1,3)\/}, i.e. for an infinitesimal Lorentz rotation of the spacetime coordinates $x^\al\lar x'^\al=x^\al+\om^\al\,_\be x^\be$ the field $\va(x)$ transforms into $\va(x)\lar \va'(x') = \va(x)+\frac{i}{2}\om^{\ga\de} \Si_{\ga\de}\,\va(x)$.

Above $\om^{\ga\de} = -\om^{\de\ga }$ are six infinitesimal constant parameters and $\Si_{\ga\de} = -\Si_{\de\ga}$ are the generators of the Lorentz algebra {\bf so(1,3)\/} in the representation space in which $\va(x)$ lives. These are hermitean and normalized to fulfil the {\bf so(1,3)\/}-Lie algebra for a generic set of Lorentz algebra generators $J_{\al\be}$
\beq  \label{1} \lbr J_{\al\be},J_{\ga\de}\rbr
= i\{\eta_{\al\ga}J_{\be\de}-\eta_{\be\ga}J_{\al\de}
+\eta_{\be\de}J_{\al\ga}-\eta_{\al\de}J_{\be\ga}\} \eeq
and hermiticity is understood w.r.t. the usual scalar product in field space.

The infinitesimal transformations above are equivalent to
\beqq x^\al\lar x'^\al&=&x^\al \\
\va(x)\lar \va'(x)&=&\Big(({\bf 1}+\Theta_\om)\, \va\Big)(x), \nonumber \eeqq
with
\beqq \Theta_\om &\equiv& -\om^{\ga\de} x_\de \pa_\ga
+\frac{i}{2}\om^{\ga\de} \Si_{\ga\de} \\
&=& \frac{i}{2}\om^{\ga\de} (L_{\ga\de} + \Si_{\ga\de}) \nonumber \eeqq
leaving spacetime coordinates unchanged whilst the Lorentz algebra element $\Theta_\om$ acts on both the spacetime and spin coordinates of the field $\va(x)$.

Rewriting the infinitesimal transformation laws for $x$ and $\va(x)$ in this equivalent way is ultimately just a choice of convention, but comes with important advantages as we shall see below and allows us to treat Lorentz transformations formally like any other inner symmetry transformation generated by a Lie group acting on the fields alone \cite{wie2}. 

Above the differential operators
\beq L_{\ga\de} = -L_{\de\ga} = -i(x_{\ga}\pa_{\de}-x_{\de}\pa_{\ga}) \eeq
are the generators of the spacetime-related part of the {\bf so(1,3)\/} transformations in field space.

Besides the generators $L_{\ga\de}$ and $\Si_{\ga\de}$ of {\bf so(1,3)\/} in field space we write
down for later use the generators $\Si_{\ga\de}^V$ of the vector representation of {\bf so(1,3)\/}
\beq \Big(\Si_{\ga\de}^V\Big)^\eta\,_\ze =
-i\big( \eta_\ga\,^\eta \eta_{\de\ze} - \eta_\de\,^\eta \eta_{\ga\ze} \big). \eeq
All these generators are normalized to obey the {\bf so(1,3)\/} algebra Eqn.(\ref{1}).

It is important to note that $\Theta_\om$ always acts upon all the spin and vector indices of the objects to its right. To illustrate the point let us look at $(\pa_\al \va)' - \pa_\al \va' = ({\bf 1}+\Theta_\om)\, \pa_\al \va - \pa_\al ({\bf 1}+\Theta_\om)\, \va$. We find
\beqq & & ({\bf 1}+\Theta_\om)\, \pa_\al - \pa_\al ({\bf 1}+\Theta_\om)\,
= \Theta_\om\, \pa_\al - \pa_\al \Theta_\om \nonumber \\
& & = \Big(-\om^{\ga\de} x_\de \pa_\ga  
+\frac{i}{2}\om^{\ga\de} \Si_{\ga\de}\Big) \pa_\al
+\frac{i}{2}\om^{\ga\de} \Big(\Si_{\ga\de}^V\Big)_\al\,^\be \pa_\be \\
& & \quad\quad -\pa_\al \Big(-\om^{\ga\de} x_\de \pa_\ga 
+ \frac{i}{2}\om^{\ga\de} \Si_{\ga\de}\Big)
= 0 \nonumber \eeqq
or $(\pa_\al \va)' = \pa_\al \va'$ as we would expect.

Next we look at the dynamics of $\va$ and assume it is determined by a Lagrangian density ${\cal L}_M(\va,\pa_\al \va)$ given in terms of the field $\va$ and its first derivatives. The field equations are then obtained by extremizing the action $S_M=\intf\,{\cal L}_M (\va,\pa_\al \va)$.

A globally Lorentz covariant Lagrangian density ${\cal L}_M(\va,\pa_\al \va)$ transforms as a scalar, i.e.
\beqq \label{7} & & {\cal L}_M(\va'(x),\pa_\al \va'(x))=
{\cal L}_M(\va(x),\pa_\al \va(x)) \\
& &\quad\quad\quad -\om^{\ga\de} x_\de \pa_\ga
{\cal L}_M(\va(x),\pa_\al \va (x)). \nonumber \eeqq
Taking into account that $\pa_\al \va' =( \pa_\al \va)'$ and that the last term in Eqn.(\ref{7}) is a pure divergence $\om^{\ga\de} x_\de \pa_\ga {\cal L}_M = \pa_\ga (\om^{\ga\de} x_\de {\cal L}_M)$ we find that the corresponding action $S_M$ is globally Lorentz invariant $S_M' = \intf\,{\cal L}_M(\va',\pa_\al \va') = S_M.$

Finally we recall that under spacetime translations $x^\al\lar x'^\al=x^\al+\ep^\al$ the field $\va(x)$ transforms into $\va(x)\lar \va'(x') = \va(x)$ with $\ep^\al$ four infinitesimal constant parameters.

The transformations above are equivalent to
\beqq x^\al\lar x'^\al&=&x^\al \\
\va(x)\lar \va'(x)&=&\Big(({\bf 1}+\Theta_\ep)\, \va\Big)(x), \nonumber \eeqq
with
\beq \Theta_\ep \equiv -\ep^\ga \pa_\ga
 \eeq
leaving the spacetime coordinates unchanged whilst the translation algebra element $\Theta_\ep$ acts on the field $\va(x)$ only as in the case of an inner symmetry transformation.

A globally Lorentz covariant Lagrangian density ${\cal L}_M(\va,\pa_\al \va)$ transforms under translations as a scalar, i.e.
\beqq \label{10}& & {\cal L}_M(\va'(x),\pa_\al \va'(x))=
{\cal L}_M(\va(x),\pa_\al \va(x)) \\
& &\quad\quad\quad -\ep^\ga \pa_\ga
{\cal L}_M(\va(x),\pa_\al \va (x)). \nonumber \eeqq
The last term in Eqn.(\ref{10}) is again a pure divergence and we find that the action $S_M$ is globally translation invariant $S_M' = \intf\,{\cal L}_M(\va',\pa_\al \va') = S_M.$

\section{Local Lorentz invariance and the gauge field $B_\al$}

\paragraph{}
In this section we gauge {\bf SO(1,3)\/}, introduce the corresponding covariant derivative and gauge field and determine its transformation behaviour.

We start extending the global to a local infinitesimal Lorentz gauge group by allowing the gauge parameters $\om^{\ga\de}$ to depend on $x$
\beqq \om^{\ga\de} &\lar& \om^{\ga\de} (x) \\
\Theta_\om = \frac{i}{2}\om^{\ga\de} (L_{\ga\de} + \Si_{\ga\de}) &\lar& \Theta_\om (x) = \frac{i}{2}\om^{\ga\de} (x) (L_{\ga\de} + \Si_{\ga\de}). \nonumber \eeqq

The infinitesimal local Lorentz transformations act in field space as
\beqq \label{12} x^\al \lar x'^\al&=&x^\al \\
\va(x)\lar \va'(x)&=&
\Bigl(({\bf 1}+\Theta_\om (x))\, \va\Bigr)(x). \nonumber \eeqq
For later use we write $\va' = \va + \de_\om \va$ introducing the variation of the field
\beqq & & \de_\om \va (x) = (\Theta_\om (x) \va)(x) \\
&=& -\om^{\ga\de}(x)  x_\de \pa_\ga \va (x) 
+\frac{i}{2}\om^{\ga\de}(x)  \Si_{\ga\de} \va (x). \nonumber \eeqq

We next require globally covariant Lagrangians ${\cal L}_M(\va,\pa_\al \va)$ to be also locally Lorentz covariant, i.e. to obey Eqn.(\ref{7}) under the local transformations Eqn.(\ref{12}). As in the Yang-Mills case this is achieved by the introduction of a covariant derivative
\beq \pa_\al\lar \nabla_\al(x)  \eeq
fulfilling
\beqq (\nabla_\al(x) \va)' &=& \nabla'_\al(x) \va' \\
\mbox{or}\quad \left({\bf 1}+\Theta_\om (x)\right)\, \nabla_\al(x)&=& \nabla'_\al(x) \,\left({\bf 1}+\Theta_\om (x)\right) \nonumber \eeqq
in analogy to $\left({\bf 1}+\Theta_\om \right)\pa_\al = \pa_\al \left({\bf 1}+\Theta_\om \right)$.

If this holds the Lagrangian transforms as a scalar as required
\beqq \label{16} & & {\cal L}_M(\va'(x),\nabla'_\al\va'(x))=
{\cal L}_M(\va(x),\nabla_{\al}\va (x)) \\
& & \quad\quad\quad -\om^{\ga\de}(x)x_\de \pa_\ga
{\cal L}_M(\va(x), \nabla_\al\va (x)). \nonumber \eeqq
This will prove sufficient to construct a locally invariant
action $S_M=\int{\cal L}_M$ further below.

Next we make the usual Ansatz for the covariant derivative
\beqq \nabla^B_\al &=& \pa_\al + B_\al(x) \nonumber \\
B_\al &=& \frac{i}{2}\Bagd(x) (L_{\ga\de} + \Si_{\ga\de}) \\
&=& -\Bagd(x) x_\de \pa_\ga
+\frac{i}{2}\Bagd(x) \Si_{\ga\de} \nonumber \eeqq
with $B_\al$ acting on the representation space of the infinitesimal Lorentz group {\bf SO(1,3)\/} in which the field $\va$ lives.

To simplify the algebra we note that we can rewrite
\beqq \nabla^B_\al &=& \pa_\al - \Bagd x_\de \pa_\ga
+\frac{i}{2}\Bagd \Si_{\ga\de} \nonumber \\
&=& \left( \eta_\al\,^\ga - \Bagd x_\de \right) \pa_\ga
+\frac{i}{2}\Bagd \Si_{\ga\de} \\
&=& d^B_\al + {\bar B}_\al \nonumber \eeqq
introducing the short-hand notations
\beq \label{19} e_\al\,^\th[B] \equiv \eta_\al\,^\th - B_\al\,^{\th\ze}x_\ze \eeq
and
\beq d^B_\al \equiv e_\al\,^\th[B]\, \pa_\th,
\quad {\bar B}_\al \equiv \frac{i}{2}\Bagd \Si_{\ga\de}. \eeq
The introduction of $e_\al\,^\th[B]$ as a book-keeping device will not only help to keep the algebra involving the gauge field components $\Bagd$ manageable, but will prove crucial to establish the equivalence of the gauge field theory presented here in terms of a dimension-one vector field to general relativity expressed in Vierbein terms. It is Eqn.(\ref{19}) on which the further relevance of the theory hinges as the "Vierbein" enters as a functional of the dynamical gauge field, and not as an independent field - a small, but crucial difference which makes the approach new and the theory potentially renormalizable and unitary at the quantum level.

Note that the explicit dependence of $e_\al\,^\th[B]$ on $x$ is linked to the $x$-dependence of $L_{\ga\de}$ and does limit permissible symmetry transformations under which $e_\al\,^\th[B] \equiv \eta_\al\,^\th - B_\al\,^{\th\ze}x_\ze$ has a properly defined transformation behaviour to global translations and global as well as local Lorentz transformations.

Next we determine the transformation law for the gauge field from
\beq \left({\bf 1}+\Theta_\om (x) \right)\, (\pa_\al + B_\al) = (\pa_\al + B_\al') \,\left({\bf 1}+\Theta_\om (x) \right) \eeq
Writing $B_\al' = B_\al + \de_\om B_\al$ we find for the variation of the gauge field
\beqq \label{22} \de_\om B_\al &=& \Theta_\om (\pa_\al + B_\al)- (\pa_\al + B_\al)\Theta_\om \\
&=& [\Theta_\om, \nabla^B_\al] = \de_\om \nabla^B_\al. \nonumber \eeqq
Hence, $B_\al$ lives in the adjoint representation of {\bf so(1,3)\/}.

The second line above can be rewritten as
\beq \Big[\Theta_\om, e_\al\,^\th[B]\, \pa_\th + \frac{i}{2} \Bagd \Si_{\ga\de}\Big] = \de_\om e_\al\,^\th[B]\, \pa_\th + \frac{i}{2} \de_\om \Bagd \Si_{\ga\de} \eeq
Equating the terms proportional to $\pa_\th$ and $\Si_{\ga\de}$ a little algebra yields
\beq \label{24}
\de_\om\Bagd= -\om^{\eta\ze}x_\ze \pa_\eta \Bagd - d^B_\al \om^{\ga\de} + \om_\al\,^\be B_\be\,^{\ga\de} + \om^\ga\,_\eta B_\al\,^{\eta\de} + \om^\de\,_\eta B_\al\,^{\ga\eta} \eeq
as well as
\beq \label{25} \de_\om\eth [B] = -\om^{\eta\ze}x_\ze \pa_\eta \eth [B]
+ e_\al\,^\ep [B] \pa_\ep (\om^{\th\ze}x_\ze) 
+\om_\al\,^\be e_\be\,^\th [B]. \eeq
Note that one can derive Eqn.(\ref{25}) directly from the transformation law Eqn.(\ref{24}) as consistency requires. 

Looking in both Eqns.(\ref{24}) and (\ref{25}) at the terms apart from the first one which is related to the local Lorentz coordinate change we find that $\Bagd$ shows the typical inhomogeneous transformation behaviour of a gauge field whilst the book-keeping device $e_\al\,^\th[B]$ transforms like a Vierbein with the lower index Lorentz rotated and the upper index multiplied by the Jacobian matrix related to the local Lorentz coordinate transformation.

We note that whilst $\eth [B]$ enters the theory like a Vierbein and transforms like one for Lorentz coordinate transformations of the specific form $x^\al\lar x'^\al=x^\al+\om^\al\,_\be(x) x^\be$, it is still a functional of $\Bagd$ and not an independent field.

Finally repeating our considerations above for global translations treated as inner symmetry transformations the variations of $\Bagd$ and $\eth [B]$ become
\beq \de_\ep \Bagd= -\ep^\ga \pa_\ga \Bagd \eeq
as well as
\beq \de_\ep \eth [B] = -\ep^\ga \pa_\ga \eth [B] \eeq
with $\ep$ constant. Note that the latter transformation takes the translation of the origin by $\ep$ properly into account for the explicitly $x$-dependent part of $\eth [B]$.

\section{The field strength tensor $G_{\al\be} [B]$ and its covariant components $\Rabgd [B]$ and $\Tabg [B]$}

\paragraph{}
In this section we introduce the covariant field strength operator $G_{\al\be} [B]$ with its covariant components $\Rabgd [B]$ and $\Tabg [B]$ and determine their transformation behaviours.

We first define the field strength tensor
\beq \label{28} G_{\al\be} [B]\equiv [ \nabla^B_\al,\nabla^B_\be] \eeq
and reexpress it as
\beqq G_{\al\be} [B]&=& [d^B_\al ,d^B_\be ] + d^B_\al {\bar B}_\be 
- d^B_\be {\bar B}_\al \\
&+& [{\bar B}_\al , {\bar B}_\be] + ( B_{\al\be}\,^\eta - B_{\be\al}\,^\eta ) \nabla^B_\eta , \nonumber \eeqq
where the last term comes from taking into account the vector character of the covariant derivative.

We further evaluate
\beq [d^B_\al ,d^B_\be ] = \left( e_\al\,^\ze [B]\,\pa_\ze e_\be\,^\eta [B]-
e_\be\,^\ze [B]\,\pa_\ze e_\al\,^\eta [B] \right)\pa_\eta \eeq

Assuming $e_\al\,^\ze [B]$ is non-singular, i.e. $\det e[B]\neq 0$ there is an inverse $e^\ga\,_\eta [B]$ with
$e^\ga\,_\eta [B]\, e_\ga\,^\ze [B]=\de_\eta\,^\ze$ and we can write
\beq [d^B_\al ,d^B_\be ] = \Habg [B]\, d^B_\ga  \eeq
introducing
\beq \Habg [B] \equiv e^\ga\,_\eta [B] \left( e_\al\,^\ze [B]\,\pa_\ze e_\be\,^\eta [B]-
e_\be\,^\ze [B]\,\pa_\ze e_\al\,^\eta [B] \right) \eeq
which is an infinite series in $B$ when explicitly expressed in terms of the dynamical gauge field. Its expansion to second order reads
\beqq \Habg [B] &=&  -B_{\al\be}\,^\ga +\, B_{\be\al}\,^\ga
- \Big(\pa_\al  B_\be\,^{\ga\de} - \pa_\be  B_\al\,^{\ga\de} \Big)\, x_\de \nonumber \\
&+& \Big( B_\al\,^{\eta\ze}\, x_\ze \pa_\eta B_\be\,^{\ga\de} 
+\, B_\eta\,^{\ga\ze}\, x_\ze \pa_\al  B_\be\,^{\eta\de} \nonumber \\
&-& B_\be\,^{\eta\ze}\, x_\ze \pa_\eta B_\al\,^{\ga\de}
-\, B_\eta\,^{\ga\ze}\, x_\ze \pa_\be  B_\al\,^{\eta\de} \\
&-& B_\al\,^{\ga\eta}\, B_{\be\eta}\,^\de +\, B_\be\,^{\ga\eta}\, B_{\al\eta}\,^\de \nonumber \\
&-& (B_{\al\be}\,^\eta - B_{\be\al}\,^\eta)\, B_\eta\,^{\ga\de} \Big)\, x_\de + O(B^3). \nonumber \eeqq
Now the necessity to work with the book-keeping device $e_\al\,^\ze [B]$ to keep the algebra manageable becomes evident. 

The introduction of $H$ allows us to recast the field strength operator in a manifestly covariant form
\beqq \label{33} G_{\al\be} [B] &=& (\Habg [B]+ B_{\al\be}\,^\ga 
- B_{\be\al}\,^\ga ) \nabla^B_\ga  \nonumber \\
&+& d^B_\al {\bar B}_\be - d^B_\be {\bar B}_\al + [{\bar B}_\al ,{\bar B}_\be] - \Habg [B]\, {\bar B}_\ga \\
&=& - \Tabg [B]\, \nabla^B_\ga + \Rab [B],  \nonumber 
\eeqq
expressing it in terms of the field strength components
\beqq \label{34} \Tabg [B] &\equiv& -(B_{\al\be}\,^\ga - B_{\be\al}\,^\ga) - \Habg [B] \\
&=& \Big( \de^\ga\,_\eta +\, e^{(1) \ga}\,_\eta [B] \Big) 
\Big( R^{(1)}_{\al\be}\,^{\eta\de} [B] +\, R^{(2)}_{\al\be}\,^{\eta\de} [B] \Big)\, x_\de +\,O(B^3)
\nonumber \eeqq
and
\beqq \Rab [B] &\equiv& \frac{i}{2} \Rabgd [B] \, \Si_{\ga\de} \nonumber \\
\Rabgd [B] &=& d^B_\al  B_\be\,^{\ga\de} - d^B_\be B_\al\,^{\ga\de}
+\, B_\al\,^{\ga\eta}\, B_{\be\eta}\,^\de  \nonumber \\ 
&-& B_\be\,^{\ga\eta}\, B_{\al\eta}\,^\de 
- H_{\al\be}\,^\eta [B]\, B_\eta\,^{\ga\de} \nonumber \\
&=&  \pa_\al  B_\be\,^{\ga\de} -\, \pa_\be  B_\al\,^{\ga\de} \\
&-& B_\al\,^{\eta\ze}\, x_\ze \pa_\eta B_\be\,^{\ga\de} +\, B_\be\,^{\eta\ze}\, x_\ze \pa_\eta B_\al\,^{\ga\de}  \nonumber \\
&+& B_\al\,^{\ga\eta}\, B_{\be\eta}\,^\de -\, B_\be\,^{\ga\eta}\, B_{\al\eta}\,^\de \nonumber \\
&+& (B_{\al\be}\,^\eta - B_{\be\al}\,^\eta)\, B_\eta\,^{\ga\de} +\, O(B^3) \nonumber \\
&=& R^{(1)}_{\al\be}\,^{\ga\de} [B] +\, R^{(2)}_{\al\be}\,^{\ga\de} [B] +\,O(B^3) \nonumber \eeqq
which are both infinite series in $B$ the expansion of which to second order we have given above. The superscript $^{(\cdot)}$ denotes the power of $B$ for a given order in the expansion. Note that in the expansion of $\Tabg [B]$ we have used the fact that
\beqq \Habg [B] &=& -\, B_{\al\be}\,^\ga +\, B_{\be\al}\,^\ga \\
&+& \Big( \de^\ga\,_\eta +\, e^{(1) \ga}\,_\eta [B] \Big) 
\Big( R^{(1)}_{\al\be}\,^{\eta\de} [B] +\, R^{(2)}_{\al\be}\,^{\eta\de} [B] \Big)\, x_\de +\,O(B^3)
\nonumber \eeqq

As much as the formal analogy to the Yang-Mills case holds when gauging the Lorentz group the appearance of the $T$-term in $G$, and $T$ and $R$ being infinite series in the gauge fields point to important additional complications in the present case.  

Let us next determine the transformation law for the field strength and its components
\beqq \label{36} \de_\om G_{\al\be} [B] &=& [ \de_\om \nabla^B_\al ,\nabla^B_\be ]
+ [\nabla^B_\al, \de_\om \nabla^B_\be ] \\
&=& [\Theta_\om, [ \nabla^B_\al,\nabla^B_\be] ] 
= [\Theta_\om, G_{\al\be} [B]] \nonumber \eeqq
using Eqn.(\ref{28}). So $G$ transforms homogenously. Rewriting above as
\beqq \de_\om G_{\al\be} [B] \!\!\!&=&\!\!\! - \de_\om\Tabg [B]\, \nabla^B_\ga 
- \Tabg [B]\, \de_\om\nabla^B_\ga + \de_\om\Rab [B] \\
= [\Theta_\om, G_{\al\be}[B] ] \!\!\!&=&\!\!\! - [\Theta_\om, \Tabg [B] ]\, \nabla^B_\ga 
- \Tabg [B]\, [\Theta_\om, \nabla^B_\ga ] 
+ [\Theta_\om, \Rab [B] ] \nonumber \eeqq
yields the homogenous transformation behaviour for $T$ and $R$
\beqq \de_\om\Tabg [B]&=& [\Theta_\om, \Tabg [B]] \\
\de_\om\Rab [B]&=& [\Theta_\om, \Rab [B]]. \nonumber \eeqq

This translates to the covariant transformation laws for the field strength components
\beqq \de_\om T_{\al\be}\,^\ga [B]&=& 
- \om^{\eta\ze}x_\ze\,\pa_\eta T_{\al\be}\,^\ga [B]
+ \om_\al\,^\eta T_{\eta\be}\,^\ga [B]\\
&+& \om_\be\,^\eta T_{\al\eta}\,^\ga [B]
+\om^\ga\,_\eta T_{\al\be}\,^\eta [B] \nonumber \eeqq
and
\beqq \de_\om \Rabgd [B] &=& - \om^{\eta\ze}x_\ze\,\pa_\eta \Rabgd [B]
+ \om_\al\,^\eta R_{\eta\be}\,^{\ga\de} [B]\\
&+& \om_\be\,^\eta R_{\al\eta}\,^{\ga\de} [B]
+ \om^\ga\,_\eta R_{\al\be}\,^{\eta\de} [B] 
+ \om^\de\,_\eta R_{\al\be}\,^{\ga\eta} [B].
\nonumber \eeqq
From Eqn.(\ref{34}) we next determine the inhomogeneous transformation law for $\Habg [B]$
\beqq \de_\om \Habg [B]&=& 
- \om^{\eta\ze}x_\ze\,\pa_\eta \Habg [B]
+ d^B_\al \om_\be\,^\ga - d^B_\be \om_\al\,^\ga \nonumber \\
&+& \om_\al\,^\eta H_{\eta\be}\,^\ga [B] + \om_\be\,^\eta H_{\al\eta}\,^\ga [B]
+\om^\ga\,_\eta H_{\al\be}\,^\eta [B]. \eeqq

For later use we finally need the change of $\dem [B]$ under an infinitesimal gauge transformation. A little algebra yields
\beq \label{42} \de_\om \dem [B]
= - \pa_\eta \left( \om^{\eta\ze} x_\ze \dem [B] \right). \eeq

\section{Underlying geometric structure}

\paragraph{}
In this section we determine the principal and the associated fibre bundles, and the connection 1- and curvature 2-form respectively to illuminate the underlying geometric structure behind our approach and compare it to the literature - using the conventions of \cite{naka}.

To illuminate the main points we trivialize the various bundles locally as we are not focusing on potentially non-trivial bundle topologies, and work with infinitesimal {\bf SO(1,3)\/} structure group transformations only.

Given that in our theory the Lorentz group acts on both the spin and spacetime coordinates of the various fields we expect some complications in comparison to the usual situation as the fibre spaces have to cater to both the spin and spacetime degrees of freedom. As it is easier to deal with these complications in the context of matter fields and the fibre bundles describing them we start with constructing such bundles first - they will turn out to be bundles associated with the {\bf SO(1,3)\/} principal bundle we will finally construct.

So let us start with two intersecting open subspaces $U_i, U_j$ of the flat Minkowski base space {\bf M\/}$^{4}$, and with fields $\va(x)$ living in a given spin $s$ re- presentation of the Lorentz group {\bf SO(1,3)\/} generated by infinitesimal transformations ${\bf 1}+\Theta_\om (x)$. Hence the fields live in a vector space of dimension $2s + 1$ which we denote by ${\bf Rep\/}_s {\bf(SO(1,3))\/}$. In bundle language the fields are nothing but local sections of the bundle we want to construct. These fields or sections may be real or complex, and are taken to be infinitely differentiable or of class $C^\infty$.

Let us take a point $x \in U_i \cap U_j$ and evaluate all scalar fields of class $C^\infty (U_i \cap U_j)$ at this point. Let us denote the resulting real or complex linear space by $C^\infty (x)$. As a consequence $C^\infty (U_i \cap U_j) =\quad \cup{\!\!\!\!\!\!\!\!\!\!\!_{_{_{x \in U_i \cap U_j}}}} C^\infty (x)$.

Now we can construct the fibre bundle {\bf E\/} corresponding to fields of spin $s$. Locally its base space is the open subspace $U_i \cap U_j \subset {\bf M\/}^{4}$, its typical fibre "attached" to a point $x \in U_i \cap U_j$ is the infinite-dimensional space $C^\infty (x)\, \times {\bf Rep\/}_s {\bf(SO(1,3))\/}$ and its structure group the Lorentz group {\bf SO(1,3)\/} acting on both the spin and spacetime degrees of freedom of the fibre space. So locally we have
\beq {\bf E\/}\sim \mbox{\bf(} U_i \cap U_j\mbox{\bf)} \times \mbox{\bf(} C^\infty (\cdot)\, \times {\bf Rep\/}_s {\bf(SO(1,3))\/}\mbox{\bf)}, \eeq
where $C^\infty (\cdot)$ denotes the typical fibre "attached" to a point $\cdot \in U_i \cap U_j$.

Note that we must have typical fibres of infinite dimension in our theory, labelled by a continuous parameter. This becomes obvious in the spin $s = 0$ case, where ${\bf Rep\/}_{\it 0} {\bf(SO(1,3))\/}$ is trivial and we would have no fibre if it were not for $C^\infty (\cdot)$. In fact, all the bundles we are concerned with are Banach bundles (see e.g. \cite{lang} on the mathematics of such bundles). Their appearance in the context of physical gauge field theories seems arguably to be novel.

Let us finally look at the transition functions $t_{ij} = {\bf 1}+\Theta_\om (x)$ which live in {\bf SO(1,3)\/}. Denoting the local trivialization of a section $t \in {\bf E\/}$ on the chart $U_i$ by $t_i(x, \va(x))$, on the chart $U_j$ by $t_j(x, \va'(x))$, then $t_j(x, \va'(x)) = t_i(x, t_{ij}(x) \va(x))$. Concretely we have 
$\va(x) \lar \va'(x) = \Bigl(({\bf 1}+\Theta_\om (x))\, \va\Bigr)(x)$ or
\beq \va'(x) = \va(x) - \om^{\ga\de}(x) x_\de \pa_\ga \va (x) 
+ \frac{i}{2}\om^{\ga\de}(x) \Si_{\ga\de} \va (x)  \eeq
which makes the action of the group element on both the spin and spacetime degrees of freedom explicit. Obviously the transition functions are nothing but local gauge transformations.

It is now easy to construct the infinite-dimensional {\bf SO(1,3)\/} principal bundle {\bf P\/} underpinning our approach. Its base space is the open subspace $U_i \cap U_j \subset {\bf M\/}^{4}$, its typical fibre "attached" to a point $x \in U_i \cap U_j$ is the infinite-dimensional space $C^\infty (x)\, \times {\bf Adj\/} {\bf(SO(1,3))\/}$ and its structure group the Lorentz group {\bf SO(1,3)\/} acting on both the spin and spacetime degrees of freedom of the fibre space. Above ${\bf Adj\/} {\bf(SO(1,3))\/}$ denotes the adjoint representation of {\bf SO(1,3)\/}. So locally we have
\beq {\bf P\/}\sim \mbox{\bf(} U_i \cap U_j \mbox{\bf)} \times \mbox{\bf(} C^\infty (\cdot)\, \times {\bf Adj\/} {\bf(SO(1,3))\/}\mbox{\bf)}. \eeq

Let us again look at the transition functions $t_{ij} = {\bf 1}+\Theta_\om (x)$ which live in {\bf SO(1,3)\/}. Denoting the local trivialization of the section $t \in {\bf P\/}$ on the chart $U_i$ by $t_i(x, {\bf 1}+\Theta_\rho (x))$, on the chart $U_j$ by $t_j(x, {\bf 1}+\Theta_{\rho'} (x))$, then $t_j(x, {\bf 1}+\Theta_{\rho'}(x)) = 
t_i(x, t_{ij}(x) ({\bf 1}+\Theta_{\rho}(x)) t_{ij}^{-1}(x))$. Concretely we have
$ {\bf 1}+\Theta_\rho (x) \lar {\bf 1}+\Theta_{\rho'} (x) 
= ({\bf 1}+\Theta_\om (x))({\bf 1}+\Theta_\rho (x))({\bf 1}-\Theta_\om (x))$ or
\beq \rho'^{\,\ga\de} (x) = \rho^{\ga\de} (x) - \om^{\eta\ze}x_\ze \pa_\eta \rho^{\ga\de}
+ \rho^{\eta\ze}x_\ze \pa_\eta \om^{\ga\de}
 + \om^\ga\,_\eta \rho^{\eta\de} + \om^\de\,_\eta \rho^{\ga\eta} \eeq
which makes the action of the group element on both the spin and spacetime degrees of freedom of an element in ${\bf Adj\/} {\bf(SO(1,3))\/}$ explicit. Again the transition functions are nothing but local gauge transformations.

Next we introduce the connection 1-form $B(x)$ on $U_i \cap U_j$
\beq B(x) \equiv B_\al (x)\, \mbox{d}x^\al = \frac{i}{2}\Bagd(x) (L_{\ga\de} + \Si_{\ga\de})\, \mbox{d}x^\al \eeq
which lives in the Lie algebra {\bf so(1,3)\/}. The compatibility condition for a transition from the chart $U_i$ to the chart $U_j$ with transition function $t_{ij} = {\bf 1}+\Theta_\om (x)$ reads
\beq B' (x) = B (x) - \mbox{d}\Theta_\om (x) + [\Theta_\om (x), B (x)] \eeq
or in components
\beq B'_\al (x) = B_\al (x) - \pa_\al \Theta_\om (x) + [\Theta_\om (x), B_\al (x)] \eeq
which is nothing but the transformation law for the gauge field $B_\al (x)$ under a local gauge transformation as stated in Eqn.(\ref{22}).

Let us turn to calculate the curvature 2-form $G (x) \equiv \frac{1}{2}\, G_{\al\be} (x) \mbox{d}x^\al\wedge \mbox{d}x^\be$ from Cartan's structure equation
\beq G (x) = \mbox{D}B (x) = \mbox{d}B (x) + B (x)\wedge B (x) \eeq 
with $\mbox{D}$ being the covariant derivative related to the exterior derivative $\mbox{d}$. In components the curvature reads
\beq G_{\al\be} (x) = \pa_\al B_\be (x) - \pa_\be B_\al (x) + [B_\al (x), B_\be (x)] = [ \nabla^B_\al,\nabla^B_\be] \eeq
which is identical to the field strength tensor as given in Eqn.(\ref{28}). Interestingly, if we re-express $G_{\al\be} [B] = - \Tabg [B]\, \nabla^B_\ga + \Rab [B] $ as in Eqn.(\ref{33}) we see $\Tabg [B]$ and $\Rabgd [B]$ emerging which look formally like the torsion and curvature tensors of a Riemann-Cartan spacetime respectively in usual differential-geometrical terms, but emerge in our context naturally as field strength components from gauging the Lorentz group on flat spacetime and - crucially - depend on the gauge field $\Bagd$ only. 

The Bianchi identity for the curvature reads 
\beqq \mbox{D}G (x) &=& \mbox{d}G (x) + B (x)\wedge G (x) -  G (x)\wedge B (x) \\
&=& \mbox{d}G (x) + [B (x), G (x)] =0. \nonumber \eeqq
Finally the compatibility condition for the curvature tensor becomes
\beq G' (x) = G (x) + [\Theta_\om (x), G (x)] \eeq
or in components
\beq G'_{\al\be} (x) = G_{\al\be} (x) + [\Theta_\om (x), G_{\al\be} (x)] \eeq
which is nothing but the homogenous transformation law for the field strength tensor $G_{\al\be} (x)$ under a local gauge transformation as stated in Eqn.(\ref{36}).

We clearly see that our theory is complementary to the gauge approaches to gravity discussed in the literature (see \cite{hehl1} for an overview) also from a geometrical point of view. The key difference is that in gauge theories of the Poincar$\acute{\mbox{e}}$ group \cite{cho2}, or more general geometric approaches to local gauge invariance \cite{chang}, spacetime as the base space becomes curved and the base space and the fibres - being finite-dimensional representation spaces of the underlying gauge group - of the corresponding bundle become "interlocked".

Specifically, in these approaches a non-trivial Vierbein or metric on the base space appears as an independent dynamical variable of mass-dimension zero - destroying the renormalizability of these theories for dimensional reasons. In our approach, an object formally looking and transforming like a Vierbein appears as well, but is in fact a functional of the basic gauge field $B$ as stated in Eqn.(\ref{19}). And more importantly, as the basic gauge field carries mass-dimension one we can construct actions which are renormalizable at least by power-counting - a task to which we turn next.

\section{Invariant actions}

\paragraph{}
In this section we determine the form of general matter actions and the most general gauge field action which are invariant under local Lorentz transformations and global translations - and having consistent field quantization in mind - which are renormalizable by power-counting.

Let us start with a globally Lorentz covariant matter Lagrangian ${\cal L}_M$. Replacing the ordinary derivatives with covariant ones in ${\cal L}_M$ and taking into account both Eqn.(\ref{16}) and Eqn.(\ref{42}) we then find for the combination $\dem [B]\,{\cal L}_M(\va,\nabla^B_\al \va)$
the transformation behaviour
\beqq & &\det e'^{-1} [B]\, {\cal L}_M(\va',\nabla'^{B'}_\al \va')=
\dem [B]\, {\cal L}_M(\va,\nabla^B_\al \va) \\ & &\quad\quad\quad
-\pa_\eta \left(\om^{\eta\ze} x_\ze \dem [B] \,
{\cal L}_M(\va,\nabla^B_\al\va)\right). \nonumber \eeqq
As a result actions of the form
\beq S_M=\intf \dem [B] \,{\cal L}_M(\va, \nabla^B_\al\va) \eeq
are locally {\bf SO(1,3)\/}-invariant. Specifically this holds true for renormalizable matter Lagrangians which are globally Lorentz covariant.

Of course, $S_M$ also remains invariant if we change from one to another inertial system by global translations or Lorentz rotations.

Next we note that the most general gauge field action invariant under local Lorentz transformations can be written as a sum of contributions with different dimensions in powers of mass.

To determine it we start by establishing the mass-dimensions of the various fields. Counting powers of mass denoted by $[.]$ we find
\beqq
& & [\pa_\al] = [\nabla^B_\al] = [d^B_\al] = [B_\al] = 1 \nonumber \\
& & [e_\al\,^\vartheta] = 0, \quad [x_\al] = -1 \\
& & [\Tabg] = [\Habg] = 1, \quad [\Rabgd] = 2. \nonumber
\eeqq

Now we can write down the most general invariant gauge field action which is renormalizable by power-counting, hence is built from the gauge fields and their first and second derivatives only. This action must be a sum of covariant Lagrangians of dimensions zero, two and four in the fields $\Bagd$ and their first and second derivatives  $\pa_\be \Bagd, \pa_\eta \pa_\be \Bagd$ respectively integrated over with $\intf \dem [B]$.

For the contribution of dimension zero in the fields we find one term
\beq S^{(0)}_G = \int {\cal L}^{(0)}_G (\Bagd)
= \La\, \intf \dem [B] \eeq
with $\La$ a constant of dimension $[\La] = 4$.

Having in mind to link our approach to General Relativity we make use of the dimensionality of Newton's gravitational constant $\it\Ga$ and recall that $\frac{1}{\ka} = \frac{1}{16\pi \it\Ga} $ has mass-dimension $[ \frac{1}{\ka} ] = 2$. We then find for the most general contribution of mass-dimension two in the fields and their first and second derivatives a sum of five terms
\beqq \label{47} S^{(2)}_G &=& \int {\cal L}^{(2)}_G 
(\Bagd, \pa_\be \Bagd, \pa_\eta \pa_\be \Bagd) \nonumber \\
&=& \frac{1}{\ka}\, \intf \dem [B]
\Big\{\al_1\, R_{\al\be}\,^{\al\be} [B]  \\
&+& \al_2\, T_{\al\be\ga} [B]\, T^{\al\be\ga} [B]
+ \al_3\, T_{\al\be\ga} [B]\, T^{\ga\be\al} [B]  \nonumber \\
&+&  \al_4\, T_\al \,^{\ga\al} [B]\, T_{\be\ga} \,^\be [B]
+  \al_5\, \nabla^B_\al T_\be \,^{\al\be} [B] \Big\}. \nonumber
\eeqq
Above $ \al_i $ are constants of dimension $[\al_i] = 0$.

Finally, for the most general contribution of dimension four in the fields there we find a proliferation of terms
\beqq S^{(4)}_G &=& \int {\cal L}^{(4)}_G 
(\Bagd, \pa_\be \Bagd, \pa_\eta \pa_\be \Bagd) \nonumber \\
&=& \intf \dem [B]
\Big\{ \be_1\, R_{\al\be}\,^{\ga\de} [B]\, R^{\al\be}\,_{\ga\de} [B] \nonumber \\
&+&  \be_2\, R_{\al\ga}\,^{\al\de} [B]\, R^{\be\ga}\,_{\be\de} [B] 
+ \be_3\, R_{\al\be}\,^{\al\be} [B] \,R_{\ga\de}\,^{\ga\de} [B] \nonumber \\
&+& \be_4\, \nabla_B^\ga \nabla^B_\de R_{\al\ga}\,^{\al\de} [B]
+ \be_5\, \nabla_B^\ga \nabla^B_\ga R_{\al\be}\,^{\al\be} [B] \nonumber \\
&+& \dots \\
&+& \ga_1\, \nabla^B_\ga T_{\al\be\de} [B]\,
\nabla_B^\ga T^{\al\be\de} [B]
+ \ga_2\, \nabla^B_\ga T_{\al\be\de} [B]\, 
\nabla_B^\ga T^{\de\be\al} [B] \nonumber \\
&+& \dots \nonumber \\
&+& \ga_j\, T^4-\mbox{terms} \nonumber \\
&+& \dots \nonumber \\
&+& \de_k\, R\,T^2-\mbox{terms},\, R\,\nabla^B\, T-\mbox{terms} \nonumber \\
&+& \dots \Big\} \nonumber 
\eeqq 
with $ \be_i $, $ \ga_j $, $ \de_k $ constants of dimension $[\be_i] = [\ga_j] = [\de_k] = 0$.

By construction
\beq \label{49} S_G = S^{(0)}_G + S^{(2)}_G+ S^{(4)}_G \eeq
is the most general action of dimension $\leq 4$ in the gauge fields $\Bagd$ and their first and second derivatives $\pa_\be \Bagd, \pa_\eta \pa_\be \Bagd$ which is locally Lorentz invariant and - having consistent field quantization in mind - renormalizable by power-counting. The actual proof of renormalizability requires the much more involved demonstration that counterterms needed to absorb infinite contributions to the perturbative expansion of the effective action of the full quantum theory are again of the form Eqn.(\ref{49}) with possibly renormalized constants.

Note that in a perturbative expansion of the quantum effective action in a dimensionless coupling $g$, with $B$ shifted as $\Bagd \ar g\, \Bagd$, terms containing second derivatives $\pa_\eta \pa_\be \Bagd$ in the gauge fields appear only in $O(g^1)$, and hence contribute to the interaction part of the Lagrangian, as the $O(g^0)$-terms containing second derivatives $\pa_\eta \pa_\be \Bagd$ are pure divergences. This is crucial and ensures that the free Lagrangian related to Eqn.(\ref{49}) contains only the gauge fields $\Bagd$ and their first derivatives $\pa_\be \Bagd$, so that any free quantum theory does not contain unphysical ghosts from higher derivative terms.

Let us take stock of where we are at this point. In analogy to the construction of Yang-Mills theories the requirement of local Lorentz invariance has led us to the introduction of a covariant derivative and a dimension-one gauge field $\Bagd$ in terms of which we are able to build locally invariant matter and gauge field actions. The additional requirement of renormalizability by power-counting limits the contributions to the gauge field Lagrangian to be of dimension $\leq 4$ in the fields and hence to a finite number of terms. Whilst - in marked complication compared to the Yang-Mills case - these terms are all infinite series in the gauge fields and their first and second derivatives at the classical level the quantized gauge field theory should be renormalizable despite the algebraic complexity at hands.

Having a viable classical and a potentially consistent quantum gauge field theory at hands the crucial question then is whether this theory can describe a fundamental force like gravitation - being in some form equivalent to General Relativity - the answer to which we turn in the remaining two sections of the paper.

\section{The associated gauge field $C_\al\,^{\ga\de} [B]$ and its field strength tensor components $\Rabgd [C]$ and $\Tabg [C]$}

\paragraph{}
In this section to prepare the demonstration of the equivalence of our theory to General Relativity we introduce the associated gauge field $C_\al\,^{\ga\de} [B]$ and its field strength tensors $\Rabgd [C[B]]$ and $\Tabg [C[B]]$ and reexpress $\Rabgd [B]$ in terms of these associated quantities.

The tensor nature of $\Tabg $ allows us to introduce a new field $C_\al\,^{\ga\de} [B]$ associated to the original gauge field $\Bagd$ by demanding that
\beq \Tabg [C] \equiv -(C_{\al\be}\,^\ga - C_{\be\al}\,^\ga) - \Habg [B] =^{\!\!\!\! !}\,\, 0 \eeq
Solving for $C_\al\,^{\be\ga} [B]$ we find
\beq \label{51} C_\al\,^{\be\ga} [B] = -\frac{1}{2}( H_\al\,^{\be\ga} [B] - H_\al\,^{\ga\be} [B] - H^{\be\ga}\,_\al [B] ). \eeq

As $\Habg [B]$ is a functional of $\eth [B]$ we find that $C_\al\,^{\ga\de} [B]$ is a functional of the original gauge field $\Bagd$ with vanishing field strength tensor components $\Tabg [C[B]]=0$.

Starting with the transformation behaviour of $H_\al\,^{\ga\de} [B]$ it is easy to show that $C_\al\,^{\ga\de} [B]$ transforms under local Lorentz transformations like the gauge field $B_\al\,^{\ga\de}$ and can be viewed as a new gauge field in its own right associated to the original one.

As a consequence we can define a new homogenously transforming tensor $K_\al\,^{\ga\de} [B]$ by
\beqq \label{52} K_\al\,^{\ga\de} [B] &\equiv& C_\al\,^{\ga\de} [B] - \Bagd \nonumber \\
\mbox{or}\quad\quad\quad\quad \Bagd &=& C_\al\,^{\ga\de} [B] - K_\al\,^{\ga\de} [B] \eeqq
which is in fact just a very complicated way to express $\Bagd$ in terms of itself. The reason to do so will become clear in the next section.

Using Eqn.(\ref{52}) we note that
\beqq \Tabg [B] &=& -(B_{\al\be}\,^\ga - B_{\be\al}\,^\ga) 
- \Habg [B] \nonumber \\
&=& -(B_{\al\be}\,^\ga - C_{ \al\be }\,^\ga [B])
+(B_{\be \al }\,^\ga - C_{\be\al}\,^\ga [B]) \\
&=& K_{\al\be}\,^\ga [B] - K_{\be\al}\,^\ga [B] \nonumber
\eeqq
which can be easily inverted
\beq \label{54} K_{\al\be\ga} [B] = \frac{1}{2}( T_{\al\be\ga} [B]
- T_{\al\ga\be} [B] - T_{\be\ga\al} [B]). \eeq
A consistency check reinserting Eqns.(\ref{51}) and (\ref{54}) into $C_\al\,^{\ga\de} [B] - K_\al\,^{\ga\de} [B]$ and using Eqn.(\ref{34}) indeed yields back $\Bagd$. We note that the raising and lowering of indices in all the expressions for $B$, $C$ and $H$ above with the Minkowski metric $\eta$ is consistent as long as we remain within the framework of our theory, i.e. its covariance w.r.t. global translations and global as well as local Lorentz rotations.

We next look at the associated covariant derivative
\beq \nabla^C_\al = d^B_\al + {\bar C}_\al [B], \quad
{\bar C}_\al [B] \equiv \frac{i}{2} C_\al\,^{\ga\de} [B] \, \Si_{\ga\de}
\eeq
and its field strength tensor and find
\beq G_{\al\be} [C] = [\nabla^C_\al, \nabla^C_\be]
= \Rab [C] \eeq
with
\beqq \Rab [C] &=& \frac{i}{2} \Rabgd [C] \, \Si_{\ga\de} \nonumber \\
\Rabgd [C] &=& d^B_\al  C_\be\,^{\ga\de}[B] 
- d^B_\be C_\al\,^{\ga\de} [B]\\
&+& C_\al\,^{\ga\eta} [B]\, C_{\be\eta}\,^\de [B]
- C_\be\,^{\ga\eta} [B]\, C_{\al\eta}\,^\de [B] \nonumber \\
&+& (C_{\al\be}\,^\eta [B] - C_{\be\al}\,^\eta [B])\, 
C_\eta\,^{\ga\de} [B]. \nonumber \eeqq

Finally a little algebra allows us to express $\Rabgd [B]$ in terms of $\Rabgd [C]$, $K_\al\,^{\ga\de} [B]$ and its covariant derivative as
\beqq \label{58}
\Rabgd [B] &=& \Rabgd [C] \nonumber \\
&-& \nabla^B_\al  K_\be\,^{\ga\de} [B] 
+ \nabla^B_\be K_\al\,^{\ga\de} [B] \\
&-& K_\al\,^{\ga\eta} [B]\, K_{\be\eta}\,^\de [B]
+ K_\be\,^{\ga\eta} [B]\, K_{\al\eta}\,^\de [B] \nonumber \\
&-& (K_{\al\be}\,^\eta [B] - K_{\be\al}\,^\eta [B])\, 
K_\eta\,^{\ga\de} [B]. \nonumber \eeqq

Here a comment is due in addition to Section 6 which clarifies the underlying geometric structure of our theory  - a comment on whether the various fields like the gauge fields $\Bagd$, $C_\al\,^{\ga\de}[B]$ together with $\eth[B]$, or the field strengths $\Rabgd[B]$ and $\Tabg[B]$ in the current approach have an intrinsic geometrical significance.

At first sight one is led to simply identify the fields above with corresponding objects of Riemannian geometry formulated in a non-coordinate basis \cite{naka}. So $\Bagd$ would then be a general connection with curvature $\Rabgd$, torsion $\Tabg$ and contortion $K_\al\,^{\ga\de}$, $\eth$ the Vierbein and $C_\al\,^{\ga\de}$ the torsion-free Levi-Civit$\grave{\mbox{a}}$ connection built from $\eth$. Technically speaking these identifications make sense as long as we remain aware that the only fundamental fields from which all these objects are built are the gauge fields $\Bagd$. All other objects are induced, most importantly the Vierbein $\eth [B]$ or the metric $g^{\eta\ze} [B] = e^{\ga\eta} [B]\, e_\ga\,^\ze [B]$ whose physical significance in terms of measuring spacetime distances in a classical context is the same as if it were an independent field, but beyond that comes with no further significance in our context.

\section{Equivalence to General Relativity}

\paragraph{}
In this section we demonstrate that the action $S^{(2)}_G$ for a specific choice of the parameters $\al_i$ depends on $\Bagd$ only through $\eth [B]$ and is the same functional of $\eth [B]$ as the Einstein-Hilbert action $S_{EH}$ is of the Vierbein $\eth$ in the tetrad formulation of General Relativity. This allows us to demonstrate the equivalence of the Lorentz gauge field theory to GR for this choice of $\al_i$ if we only take into account $S^{(0)}_G, S^{(2)}_G$ and scalar matter.

We start by twice contracting Eqn.(\ref{58}) with $\eta$ to reexpress $R_{\al\be}\,^{\al\be} [B]$
\beqq 
R_{\al\be}\,^{\al\be} [B] &=& R_{\al\be}\,^{\al\be} [C[e[B]]] \nonumber \\
&-& \nabla^B_\al  K_\be\,^{\al\be} [B] 
+ \nabla^B_\be K_\al\,^{\al\be} [B] \\
&-& K_\al\,^{\al\eta} [B]\, K_{\be\eta}\,^\be [B]
+ K_\be\,^{\al\eta} [B]\, K_{\al\eta}\,^\be [B] \nonumber \\
&-& (K_{\al\be}\,^\eta [B] - K_{\be\al}\,^\eta [B])\, 
K_\eta\,^{\al\be} [B]. \nonumber \eeqq
Using Eqn.(\ref{54}) to rewrite $K_\al\,^{\ga\de} [B]$ in terms of $\Tabg [B]$
we find after a little algebra
\beqq 
R_{\al\be}\,^{\al\be} [B] &=& R_{\al\be}\,^{\al\be} 
[C[e[B]]] \nonumber \\
&-& 2 \nabla^B_\al  T_\be\,^{\al\be} [B] 
+ T_\al\,^{\eta\al} [B]\, T_{\be\eta}\,^\be [B] \\
&+& \frac{1}{2} T_{\al\be\ga} [B]\, T^{\ga\be\al} [B] 
+ \frac{1}{4} T_{\al\be\ga} [B]\, T^{\al\be\ga} [B],
\nonumber \eeqq
where we explicitly reemphasize the complicated functional dependence of $C_\al\,^{\ga\de} [e[B]]$ on $\Bagd$ through $\eth [B]$.

Inserting this expression into Eqn.(\ref{47}) we can rewrite $S^{(2)}_G$ as
\beqq S^{(2)}_G &=& \frac{1}{\ka}\, \intf \dem [B]
\Big\{ \al_1\, R_{\al\be}\,^{\al\be} [C[e[B]]] \nonumber \\
&+&  \left( \al_2 + \frac{\al_1}{4}\right)\,
T_{\al\be\ga} [B]\, T^{\al\be\ga} [B] \\
&+& \left( \al_3 + \frac{\al_1}{2}\right)\,
T_{\al\be\ga} [B]\, T^{\ga\be\al} [B] \\
&+& \left( \al_4 + \al_1 \right)\,
T_\al \,^{\ga\al} [B]\, T_{\be\ga} \,^\be [B] \nonumber \\
&+& \left( \al_5 - 2 \al_1\right)\, \nabla^B_\al T_\be \,^{\al\be} [B] \Big\} \nonumber
\eeqq
which is a functional of the gauge fields $\Bagd$ and its first and second derivatives $\pa_\be \Bagd, \pa_\eta \pa_\be \Bagd$ only, albeit a very convoluted one.

Now the reason for introducing the associated gauge fields $C_\al\,^{\ga\de} [e[B]]$ and for re-expressing $\Rabgd [B]$ through $\Rabgd [C[e[B]]]$ plus $K$-terms in the preceeding section becomes clear.

Setting
\beq \label{63} \al_1 = 1,\quad \al_2 = -\frac{1}{4},\quad \al_3 = -\frac{1}{2},\quad
\al_4 = -1,\quad \al_5 = 2
\eeq
$S^{(2)}_G$ reduces to
\beq S^{(2)}_G = \frac{1}{\ka}\, \intf \dem [B] \,
R_{\al\be}\,^{\al\be} [C[e[B]]] \eeq
which is a functional of $\Bagd$ and $\pa_\be \Bagd$ through $\eth [B]$  and its first derivatives only.

Crucially, $S^{(2)}_G$ is by construction the same functional of $\eth [B]$ as the Einstein-Hilbert action $S_{EH}$ is of the Vierbein $\eth$ in the tetrad formulation of General Relativity.

The same argument trivially applies to $S^{(0)}_G$ and the cosmological constant contribution to GR and to actions describing purely scalar matter.

Hence, if we look at the truncated Lorentz gauge field theory taking into account only $S^{(0)}_G$ and $S^{(2)}_G$ and restrict ourselves to scalar fields for the matter contribution the resulting theory is in effect equivalent to GR with a cosmological constant term, coupled to scalar matter.

Formally this becomes evident looking at a scalar field minimally coupled to $\Bagd$ which is described by the Lagrangian ${\cal L}_M = {\cal L}_M (\va, d^B_\al \va)$.  Adding the locally Lorentz invariant gauge field action $S^{(0)}_G [e[B]] + S^{(2)}_G [e[B]]$ the total action then is a functional of $\Bagd$ and its first derivatives through $\eth [B]$ and its first derivatives only
\beqq S &=& S^{(0)}_G + S^{(2)}_G + S_M \nonumber \\
&=& \intf \dem [B] \Big\{ \Lambda + \frac{1}{\ka}\, R_{\al\be}\,^{\al\be} [C[e[B]]]  \\
&+&  {\cal L}_M (\va, \eth[B] \pa_\th \va) \Big\}. \nonumber
\eeqq
Evoking the chain rule the field equations for $\Bagd$
\beq
\frac{\de S}{\de \Bagd} = \int \dem [B]
\frac{\de S}{\de \eth [B]}\, \frac{\de \eth [B]}{\de \Bagd} = 0
\eeq
are equivalent to
\beq \label{67}
\frac{\de S}{\de \eth} = 0
\eeq
as by construction there are no terms in the action $S$ which depend on $\Bagd$ not through $\eth [B]$. The latter equations can be viewed as field equations for $\eth$ in their own right and have the same functional dependence on $\eth [B]$ as have the Einstein equations on the Vierbein $\eth$ in the tetrad formulation of General Relativity. However, we note that here the underlying fundamental fields are still the $\Bagd$ which, however, have been completely shielded away.

So whilst Eqn.(\ref{67}) looks formally the same as the Einstein equations for the Vierbein and allow for solving for $\eth$ the former still carries a memory of having emerged from an underlying dimension-one vector gauge field theory which becomes immediately explicit when taking into account terms in the full action which depend directly on $\Bagd$, e.g. from $S^{(4)}_G$ or from including matter with spin.

We might wonder at this point what the conditions on the various para- meters in $S_G$ are such that gravitation at the classical level can be described by an $S_G[B]$ which is a functional of  $\eth[B]$ in essence - hence with terms with an explicit dependence on $\Bagd$ massively suppressed. The two conditions are (1) that all the parameters in $S^{(4)}_G$ are suppressed as compared to the parameters in $S^{(2)}_G$ - which Nature ensures by $\kappa$ being small - and (2) that the parameters $\al_i$ above are close to their values as in Eqns.(\ref{63}).

\section{Conclusions}

\paragraph{}
In this paper we have developed in the first part a consistent gauge theory of the Lorentz group in terms of one gauge field of mass-dimension one. In the second part we then have demonstrated that in a certain limit the theory presented is equivalent to General Relativity as expressed in Vierbein terms making it a candidate gauge theory to describe gravitation.

At first sight the theory presented could easily be misread as General Relativity  - or a generalization of it - expressed in anholonomic coordinates \cite{cho1}, or as a gauge theory of the Poincar$\acute{\mbox{e}}$ group \cite{cho2}. We recall that in both those approaches the dynamical field variables are the Vierbein $e$ and an affine connection $B$, which in the limiting case of vanishing torsion can be expressed in terms of $e$ alone. Both theories are built to be invariant under general coordinate transformations in addition to local Lorentz rotations acting on spin-degrees of freedom.

The key point of the theory presented here is that it depends on one gauge field only. All objects needed to formulate the theory are expressed in terms of that one gauge field, most notably an expression introduced to simplify the sometimes complex algebra which looks and transforms like a Vierbein. The price to pay is that the theory is invariant under local Lorentz transformations only, and not under general coordinate transformations. All this points to subtle, yet fundamental differences to the existing approaches which might allow to both overcome the obstacles to consistent quantization as well as to add to a refined understanding of gravity at the classical level. It is these subtle differences which make the approach new. 

At the classical level it is reassuring that with the appropriate choice of constants $\al_1$ to $\al_5$ as in Eqns.(\ref{63}) the leading terms $S^{(0)}_G + S^{(2)}_G$ in the presence of scalar matter only reduce to a theory expressible in terms of $e[B]$ only which features an additional accidental symmetry under general coordinate transformations, and is completely equivalent to Einstein's theory of general relativity. Hence, it reproduces all the classical GR results e.g. on redshift, on the bending of light, on the perihelion precession of Mercury or on frame dragging. As $S^{(0)}_G + S^{(2)}_G$ represents the leading terms in a low energy or large distance "expansion" (see \cite{don}) we expect also other key results of general relativity to hold at low energies or large distances - the term $S^{(4)}_G$ becomes relevant only at distances of the order $\sqrt{\ka}$ which is the Planck length, where the theory will completely change its behaviour and in essence becomes of quantum nature.

In addition, the explanation of phenomena such as the accelerated expansion of the universe or the galaxy rotation curves not compatible with the observed matter distribution might after all not be linked to dark energy or dark matter, but rather to a refined theory of gravitation.

At the quantum level consistent quantization has faltered in essence due to the fact that one indispensable dynamical field in the traditional approaches, $g$ or $e$, carries dimension zero which renders the underlying theories non-renormalizable or non-unitary.

In the gauge theory of the Lorentz group presented here the only gauge field has mass-dimension one. This has allowed us to construct the most general action for that field which is renormalizable by power-counting - potentially opening the route to a theory which is both renormalizable and unitary. To make further progress a full renormalization proof is required as well as the demonstration that the physical S-matrix is unitary. The latter is complicated by the fact that {\bf SO(1,3)\/} is not compact. Hence, one has to show that the canonical quantization of both the free and interacting gauge fields allows for the definition of positive-energy, positive-norm states and a corresponding relativistically-invariant physical Fock space for these fields - and that negative-norm states completely decouple. As this is related to the non-compactness of the gauge group and not to the occurence of higher derivatives of the gauge field progress should be possible.

\end{document}